%% file: Main.tex
\documentclass{article}

\usepackage{amsmath,amsfonts}
\usepackage{algorithmic}
\usepackage{algorithm}
\usepackage{array}
\usepackage{textcomp}
\usepackage{stfloats}
\usepackage{xcolor}
\usepackage{url}
\usepackage{verbatim}
\usepackage{graphicx}
\usepackage[skip=0.333\baselineskip]{caption}
\usepackage{subcaption}
\usepackage{booktabs}
\usepackage{multirow}
\usepackage{authblk}
\usepackage[square,numbers]{natbib}
\usepackage{xurl}

\begin{document}

\title{Shortlisting Protection Configurations for HVDC Grids and Electrical Energy Hubs}

\author[1,2]{Merijn Van Deyck}
\author[1,2]{Geraint Chaffey}
\author[1,2]{Dirk Van Hertem}

\affil[1]{{Department of Electrical Engineering, KU Leuven, Leuven, Belgium} \newline}
\affil[2]{{Etch-EnergyVille, Genk, Belgium}}

\date{}

\maketitle

\begin{abstract}
\input{0-Abstract}
\end{abstract}

\textbf{Keywords:}
HVDC grid protection, Electrical energy hubs, Energy islands, HVDC substation planning

\input{1-Introduction}

\input{2-Methodology}

\input{3-Results}
\input{4-Conclusion}

\section*{Acknowledgement}
This work received funding from the DIRECTIONS project, supported by the Energy Transition Fund, DG Economy, Belgium and was partially supported by the project `Innovative solutions for underground high-voltage lines and grids', funded by the Flemish Government. The authors would like to thank Prof. Dr. ir. Willem Leterme for the initial brainstorm and discussions.


\bibliography{References.bib}

\end{document}

%% file: 0-Abstract.tex
This paper proposes a methodology for shortlisting protection system configurations for large HVDC switching stations, which are expected in multiterminal HVDC grids and electrical energy hubs (or \textit{energy islands}). This novel approach focuses on the configuration of protection equipment and the arrangement of lines and converters in various protection zones, instead of expert decisions on protection strategies based on numerous simulations. 
A graph-based approach that allows high-level evaluation of possible DC fault impacts is presented. This fault impact evaluation method can evaluate many possible protection configurations allowing the selection of less obvious choices, as experts cannot consider all possible configurations, especially when the switching station size increases. A filtering process is applied to reduce the number of possible configurations based on multiple protection performance metrics which are evaluated for different power flow scenarios. The results for these performance metrics can be compared for configurations with different numbers of HVDC circuit breakers to assess the benefit of increasing the amount of protection equipment in different network topologies. It is also shown that, through continued filtering using additional performance metrics or fault scenarios, the number of possible breaker, cable and converter configurations can be further reduced, leading to a protection design that is well suited for many operational scenarios. The results of the shortlisting process provide insights on the required number of HVDC circuit breakers to limit fault impacts to a given value. Moreover, observed trends in the results could, in future studies, contribute to new design principles and priorities, allowing system developers to more effectively design HVDC protection systems for different operational scenarios and possible investment levels.

%% file: 1-Introduction.tex
\section{Introduction}
\label{sec:introduction}
Recent declarations by surrounding countries aim to transform the North Sea into \textit{Europe's green power plant}~\cite{OstendDeclaration}. To achieve this, these countries aim to increase offshore wind capacity to 300~GW by 2050. Significant transmission infrastructure investments will be required, both onshore and offshore, to integrate these vast amounts of offshore energy into the existing power systems. 

The prospective development of offshore multiterminal HVDC (MTDC) grids offers flexible transmission capacity, well suited for high power capacities and long distances~\cite{bookDirk}. The electrical energy hub concept - commonly also known as \textit{energy islands} - is expected to become an important aspect of these MTDC grids in the future. In these hubs, offshore equipment can be centralized to serve both as a central node in interconnected MTDC grids, and for the large-scale integration of renewable energy sources. Multi-GW offshore energy hubs are under development in Belgium~\cite{Elia2023Website} and Denmark~\cite{DanishWebsite, Bornholm} and have been considered for several other countries~\cite{EsbjergCooperation}.

The protection of MTDC grids has been a key obstacle to the realization of these systems~\cite{Dirk2010DCgrids}. The choice of a \textit{protection strategy} is a key step in the planning of future MTDC grids~\cite{leterme2019designing, PROMOTioN_D4_2}. These strategies (fully selective, non-selective, and partially selective protection) dictate the selectivity of the protection system and consequently have a large influence on the protection equipment requirements and the impact of DC faults.  

\subsection{Existing protection design methodologies}
HVDC circuit breakers (DCCB) are significantly more complex and expensive than AC breakers. Therefore, several methodologies exist for assessing the need for these breakers in MTDC grids. In~\cite{PROMOTioN_D4_2, leterme2019designing}, different implementations of the fundamental protection strategies are considered for a four terminal benchmark network. A comparison of these strategies and implementations is performed based on protection performance metrics such as component count, AC grid impact, risk analysis and extensibility. Time-domain electromagnetic transient (EMT) simulations generate a thorough understanding of the transient fault response related to each strategy implementation. Though this method allows a detailed comparison of possible protection implementations for an MTDC grid, it relies on expert-based selections of the possible protection strategies and implementations. The use of EMT studies to compare these options brings little generic insights and imposes a high computational load on the design, which will significantly increase when the network size increases~\cite{subedi2021review}.

A protection design approach for a larger MTDC grid that consists of multiple interconnected offshore electrical energy hubs is presented in~\cite{NSWPH_II, SuperGridInstitute}. The protection design is carried out in a modular way based on a cost-benefit analysis methodology from~\cite{PROMOTioN_D4_7}. A standardized DC-side busbar topology and protection strategy are determined based on estimated costs, AC grid constraints and fault ride through capabilities of offshore wind farms. Load flow studies and EMT simulations are performed to dimension the DC reactors.  The protection design is standardized such that it can be uniformly applied, leading to an implementation of the fully selective strategy in each hub. This modular approach is time efficient for designing the protection system for several electrical energy hubs, since the cost-benefit analysis and component sizing is only performed once. However it is unclear whether the chosen protection configuration is actually the ideal option for every hub. It seems likely that better options may exist for specific hubs and that these may vary since different AC grid requirements, fault ride through capabilities or energy hub sizes may apply across the system. Therefore, a protection design approach that facilitates the efficient design of several hubs while considering the specific context of each system may improve the overall HVDC protection design. 

Methodologies for HVDC protection design based primarily on AC grid frequency stability limits have been proposed in~\cite{DCcont2017KUL, BrantlLjubljana2021,Dave2022Thesis}. These methods evaluate protection strategies by considering the loss of power transfer capacity during the fault clearing process. Depending on the duration of this power flow interruption, the frequency change in the AC grid can be calculated. By taking into account frequency stability limits, this results in strict limitations of the protection zone size, which follows from the protection strategy choice, and of the operating time of the DC protection and grid restoration sequence. In~\cite{Dave2022Thesis}, this method is used for a techno-economic evaluation of the non-selective and fully selective strategies based on the expected costs for protection equipment and frequency containment reserves required to maintain AC grid stability. The evaluation of loss of power transfer capacity is an efficient method to evaluate possible DC fault impacts that does not require detailed time-domain simulations. As a result, many possible designs, operational scenarios or network configurations can be considered in this way. 

These methodologies focus on applying a fundamental protection strategy to a predefined HVDC grid topology. However, the system-level application of a protection strategy does not inherently describe the full implementation of the protection system. For example, the partially selective protection strategy implies that part of the MTDC grid remains in continuous operation while another part, containing the faulted element, is de-energized as the fault is cleared~\cite{PROMOTioN_D4_2}. This principle can be implemented in several ways in heavily meshed grids or radial grids with many connections. Moreover, other configurations that are not fully described by fundamental protection strategies, e.g. preventive grid reconfiguration~\cite{Preventive_Decoupling_paper}, can be used together with other protection strategies to improve the effectiveness of the DC protection system, meaning that low fault impacts are achieved for low protection system costs. 

\subsection{Contribution and outline}
The arrangement of different connections, and particularly the inclusion of DCCBs between these connections, is expected to become an important factor in the DC protection design of future MTDC grids with large HVDC switching stations and electrical energy hubs. This design of the protection \textit{configuration} encompasses more than is typically described by fundamental protection strategies and leads to a complex design problem with a large number of permutations. However, this design method offers a more open approach to HVDC protection design. It ensures that installed equipment is efficiently utilized and over-investments are avoided.

This approach to protection configuration design introduces many possible options, especially as the number of lines, converters and DCCBs increase. Consequently, performing detailed time-domain studies or techno-economic cost-benefit analyses for each configuration would not be possible since, depending on the network size, millions of configurations would have to be evaluated. Therefore, this paper proposes a shortlisting approach that selects suitable HVDC switching station configurations based on high-level indicators related to possible DC fault impacts. After this initial shortlisting, detailed studies can be performed to complete the full substation and protection design. With this approach, MTDC grid developers can design protection systems with a focus on specific switching stations, rather than applying grid-level strategies. By focusing on the high-level configuration in this initial stage, it also becomes easier to perform the (co-)design of large systems consisting of several hubs. The full shortlisting methodology is explained in Section~\ref{sec:methodology}. In Section~\ref{sec:results}, the method is applied to three test cases, which represent electrical energy hubs of different sizes.

%% file: 2-Methodology.tex
\section{Energy hub protection configuration shortlisting methodology}
\label{sec:methodology}

This section describes the models and methods used for shortlisting protection configurations. A graph-based representation of electrical energy hubs that allows efficient fault impact calculations is proposed. The principles of the fault clearing method are then explained, followed by the key assumptions of the method. The shortlisting process is then described generally, after which a generic electrical energy hub test case is used to illustrate the method in the final subsection. 

\subsection{Modeling approach}
The arrangement of protection zones in an MTDC grid is crucial to the HVDC protection design. These zones are bordered by elements that can interrupt DC fault currents, e.g. DCCBs or fault-blocking converters. Without these fault-blocking components, DC fault currents cannot be interrupted by typical AC switchgear due to their lack of natural zero crossings. Consequently, if a fault occurs within a protection zone, all elements in this zone must be de-energized for a short time, allowing the faulted element to be cleared at near-zero current by mechanical switchgear such as high-speed disconnectors~\cite{SGI_non_selective}. In typical HVDC protection designs, the size of the protection zones, and the number of elements which need to be de-energized during the fault clearance, are determined by the protection strategy, i.e. the entire HVDC grid for the non-selective strategy, part of the grid in the partially selective strategy and only the faulted element in the fully selective strategy~\cite{leterme2019designing}. By allowing healthy parts of the grid to remain in continuous operation while the fault is cleared, selective protection strategies reduce the momentary loss of active power infeed into AC grids, limiting the risk of frequency deviations in these grids~\cite{DCcont2017KUL, BrantlLjubljana2021}. 

HVDC protection designs can be evaluated based on the protection zones resulting from the configuration of converters, cables and DCCBs and by evaluating possible fault impacts. These can be quantified by the \textit{worst-case loss of power transfer}. This metric ($P_{LoI}$ in Eq.~\ref{eq:P_impact}) is defined as the maximum loss of infeed (LoI), i.e. the difference in power flow before and during the fault ($P_{PreFault}$ and $P_{Fault}$), in each connected AC grid ($G$), caused by a fault in any protection zone ($z$).
\begin{equation}
\label{eq:P_impact}
    P_{LoI} = \max_{z}\left(\sum_{G}P_{PreFault,G}-P_{Fault,G,z}\right)
\end{equation}

The worst-case loss of power transfer metric is related to the AC grid frequency stability, but does not require detailed knowledge of the AC grid inertia, grid loading and configuration, or of the DC protection operating time. Consequently, protection configurations can be compared in a simplified manner. Since detailed grid models are not required for this, a graph-based representation of the HVDC and AC grids is used to represent the configuration of the grid before and during the faulted state.

\begin{figure}[htbp]
\begin{subfigure}{.47\textwidth}
  \includegraphics[width=\linewidth]{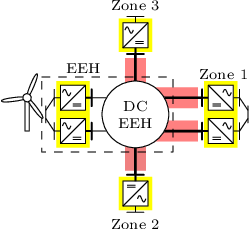}
  \caption{System-level diagram}
  \label{fig:small_case_SLD}
\end{subfigure}\hfill
\begin{subfigure}{.47\textwidth}
  \includegraphics[width=\linewidth]{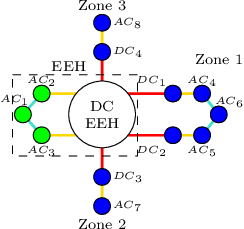}
  \caption{Graph-based representation}
  \label{fig:SmallGraph}
\end{subfigure}\hfill
\caption{Per-pole representation of an example electrical energy hub test case representing near-future designs.}
\label{fig:small_case_representation}
\end{figure}

Fig.~\ref{fig:small_case_representation} shows an example electrical energy hub test case represented as a system-level diagram in Fig.~\ref{fig:small_case_SLD} and using a graph-based representation in Fig.~\ref{fig:SmallGraph}. This example network consists of an electrical energy hub (EEH), indicated by the dashed box, with two parallel converter stations and four connections to three different synchronous AC zones. In both diagrams, converters and HVDC cables are indicated in yellow and red respectively. This network topology resembles early energy hub designs in Belgium and Denmark and consequently serves as a representation for near-future electrical energy hubs. Since, further in the future, electrical energy hubs are expected to increase in size, this example is termed the \textit{small test case}. A \textit{medium-} and \textit{large test case} are introduced for additional analysis in Section~\ref{sec:results}. Although a per-pole representation is used here, any configuration, e.g. monopolar or bipolar, can be used in the actual design. The case studies in this paper consider a bipolar converter configuration.

The proposed network model represents key elements in the grid, as relevant to protection studies, i.e. cables, converters and switchgear (blue lines in Fig.~\ref{fig:SmallGraph}) as edges between graph nodes. Electrical network configurations can be varied by changing the nodes at which each edge is connected. In the diagrams in Fig.~\ref{fig:small_case_representation}, the DC side of the electrical energy hub (DC EEH) is not explicitly depicted since the design of this area will be determined by the protection configuration design method.

Nodes and edges are characterized as \textit{internal}, i.e. on the electrical energy hub (green nodes in Fig.~\ref{fig:SmallGraph}), or \textit{external} (blue nodes in Fig.~\ref{fig:SmallGraph}) and as AC or DC, i.e. connected to the AC or DC side of the network. 

A \textit{fault-blocking} characteristic is assigned to edges in the graph model that represent circuit breakers or other fault blocking elements, indicating that fault currents can be interrupted there. This allows the calculation of \textit{protection zones} based on the location of these fault-blocking edges. Consequently, any configuration of protection zones can be achieved by varying the location of these fault-blocking edges and the connections of converter and cable edges to the DC nodes in the electrical energy hub. 

External AC nodes represent different synchronous systems connected to the electrical energy hub. In future systems, electrical energy hubs may be connected to a larger multiterminal HVDC grid containing additional electrical energy hubs. Though the examples illustrated here focus on a single electrical energy hub, these larger HVDC grids with multiple energy hubs can be considered using the proposed shortlisting method.

These simplified representations of HVDC- and AC grids and of the HVDC protection designs allow the creation and evaluation of large numbers of electrical energy hub topologies and protection implementations. Therefore, though this modeling approach does not give the full detail of the technical implications of grid topologies and protection designs, it is effective for shortlisting. Consequently, this shortlisting step serves as a precursor to more detailed protection design studies. In these studies, selected configurations and topologies can subsequently be used to perform detailed design and analysis in e.g. EMT simulation.

\subsection{Fault impact evaluation}
\begin{figure}[htbp]
\begin{subfigure}{.4\textwidth}
 \includegraphics[width=\linewidth]{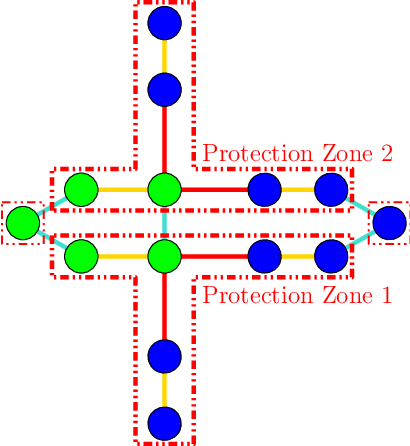}
  \caption{Definition of protection zones based on fault-blocking edges}
  \label{fig:sequence_zones}
\end{subfigure}\hfill 
\begin{subfigure}{.4\textwidth}
  \includegraphics[width=\linewidth]{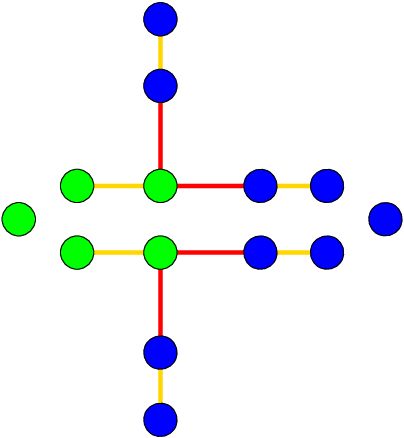}
  \caption{Protection zones split into sub-graphs}
  \label{fig:sequence_subgraphs}
\end{subfigure}

\medskip 
\begin{subfigure}{.4\textwidth}
  \includegraphics[width=\textwidth]{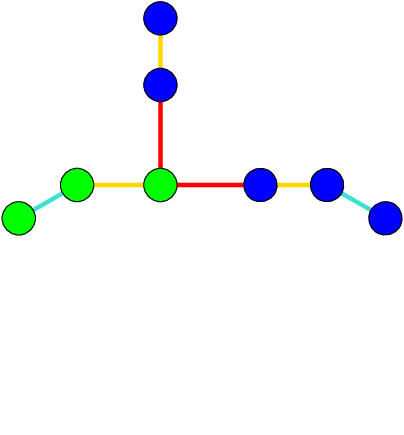}
  \caption{Network during the faulted state}
  \label{fig:sequence_fault_state}
\end{subfigure}\hfill 
\begin{subfigure}{.4\textwidth}
  \includegraphics[width=\linewidth]{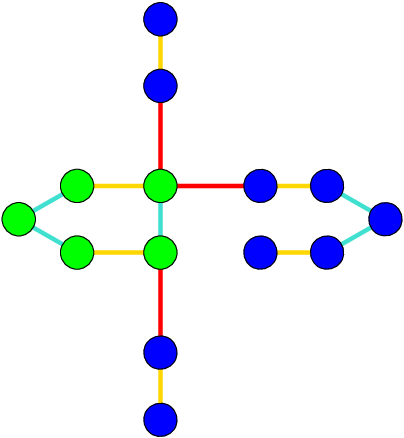}
  \caption{Network during the post-fault state}
  \label{fig:sequence_post_fault}
\end{subfigure}
\caption{Grid representation for the evaluation of fault impacts.}
\label{fig:impact_sequence}
\end{figure}

The graph-based network modeling approach is used to compare grid topologies and protection designs by evaluating worst-case fault impacts for different DC protection configurations. By evaluating the fault-blocking characteristics of the edges in the graph model, the network can be split into sub-graphs representing the protection zones, as depicted in Fig.~\ref{fig:sequence_zones}-\ref{fig:sequence_subgraphs} where an example configuration with one DCCB is used. HVDC protection operates such that the full protection zone is de-energized if any element within that zone is faulted. Therefore, the network can be represented in the faulted state by removing the sub-graph containing the faulted element from the pre-fault graph as exemplified in Fig.~\ref{fig:sequence_fault_state} for an example fault anywhere in Protection Zone 1. In the post-fault state, it is assumed that the faulted element is removed from the system and the remaining elements in the grid are re-energized and reconnected to the grid, as represented in Fig.~\ref{fig:sequence_post_fault} for an example cable fault.

\begin{figure}[htbp]
    \centering
    \includegraphics[width=\linewidth]{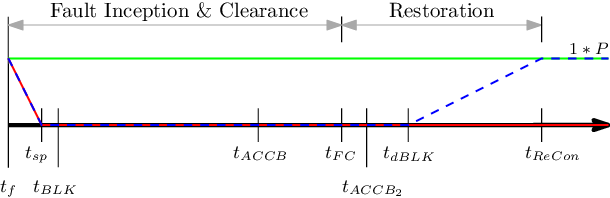}
    \caption{Availability of controlled active power during the fault clearing and restoration process on a cable in the healthy protection zone (green), on a healthy cable in the faulted protection zone (blue, dashed) and on the faulted cable (red).}
    \label{fig:time_domain_graph}
\end{figure}

Fig.~\ref{fig:time_domain_graph} illustrates an example time-domain representation of the fault clearing process. The figure shows the available controlled active power per cable, out of the maximum cable rating $P$, during the fault clearing process for a cable in a healthy protection zone (green), a healthy cable in the faulted protection zone (blue), and the faulted cable (red). The faulted state of the network lasts from the opening of the DCCBs and the splitting of the grid into healthy and unhealthy protection zones at $t_{sp}$ until the reconnection of healthy grid elements at $t_{ReCon}$ which follows the fault clearing and converter deblocking. Though the exact duration of this fault clearing process depends on many factors such as the exact fault location and grid configuration, it can be estimated that the grid operates in the faulted state for up to several seconds~\cite{DantasProgressiveIsolation2018}.

Using this modeling approach for faults, their impact is evaluated by calculating the loss of active power transfer between the electrical energy hub and onshore AC grids in the pre-fault and faulted states, as represented by Eq.~\ref{eq:P_impact}. The electrical characteristics of the grid, i.e. the voltages at each node and the impedance of the branches, are not included in the models. Therefore, a different approach is used to model the power flows. This approach relies on the \textit{capacity} and \textit{length} characteristics of graph edges and on predefined infeed and demand capacities at the AC nodes\footnote{More complex power flow scenarios can be implemented if additional offshore wind farms or connected electrical energy hubs are considered, i.e. by defining additional infeed capacity at external DC nodes. This may lead to a more accurate representation of the expected power flows in the grid, though this evaluation is left out of the scope of this conceptual paper.}. To ensure that the considered faults lead to the highest loss of power transfer, and thus represent worst-case fault impacts, the grid is assumed to be maximally loaded. Any operational scenario where the grid is operated below the maximum rating would consequently lead to a lower total LoI and lower total fault impacts. Different power flow scenarios are created by changing the direction of power flow to and from the different AC nodes. By evaluating different power flow scenarios, it can be ensured that all relevant fault cases are considered and that the selected protection configuration operates effectively in each.

In the illustrative examples in this paper, it is assumed that all cables and converters have the same capacity of $1P$, e.g. 1~GW (per pole in a bipolar system). The loss of power transfer in different fault scenarios and protection configurations can therefore be expressed in terms of multiples of $P$, making the method applicable for systems of different ratings and in different configurations. Mixed systems, i.e. systems with multiple ratings or combinations of monopolar and bipolar systems with or without a dedicated metallic return can also be considered. In rigid bipolar systems with no metallic return, a pole-to-ground cable fault results in a $2P$ loss of power transfer capacity, since both poles are affected. Similarly, pole-to-pole faults cause twice the impact of pole-to-ground faults in bipolar grids with a dedicated metallic return. By integrating these aspects in the protection shortlisting methodology, complex systems can be evaluated. However, in this paper, the illustration of the method on the example test cases is limited to the single-pole representation of a system where each outage causes a permanent capacity reduction of $1P$.

Under these assumptions, the demand and infeed settings for the example grid test case can, for example, be set as listed in Tab.~\ref{tab:DemandInfeed}. In specific grid topologies where the total transmission capacity exceeds the infeed capacity on the electrical energy hub, the grid can be fully loaded by allowing onshore zones to export additional power. The onshore zone fulfilling this role can be varied to achieve different power flow scenarios that each fully load the HVDC grid. 

\begin{table}[htbp]
    \centering
    \caption{Example infeed and demand settings in the small test case example, leading to a maximal grid loading.}
    \label{tab:DemandInfeed}
    \begin{tabular}{cc}
    \toprule
        Node & Infeed ($+$) / Demand ($-$)\\ \midrule
        $AC_1$ & $+2P$\\
         $AC_6$& $-2P$ \\
         $AC_7$& $+1P$ \\
         $AC_{9}$& $-1P$ \\ \bottomrule
    \end{tabular}
\end{table}

Given a set of power infeed and demand settings, a \mbox{graph theory-based} shortest path algorithm that is adapted to include available line capacity is used to match demand and infeed nodes and thus create indicative possible power flow scenarios for the fault impact analyses. Power flows are created in this way for the pre-fault and faulted network. In the faulted state, the available transmission capacity is reduced, which means that not all demand in the AC grid nodes can be satisfied, causing a momentary loss of infeed. By calculating this LoI in different electrical energy hub configurations, the efficiency of the protection design is evaluated. 
 
\subsection{Model assumptions}
The representation of AC and HVDC grids, faults, and the fault clearing process described in the previous subsections, offers a simplified method to evaluate fault impacts with enough detail to compare different electrical energy hub protection designs based on the configuration of the protection zones. However, this simplified representation makes some technical assumptions of which the most important are summarized in this subsection. The succession of pre-fault, faulted and post-fault states assumes a protection system that can detect, locate, and interrupt fault currents. In addition, it is assumed that the grid elements outside the faulted protection zone remain in continuous operation, allowing the power flow to persist in the remaining grid elements in the faulted-state network. This assumption may influence some detailed aspects of the protection system implementation, such as converter fault ride through requirements and DC protection coordination, that do not typically need to be considered at this stage of the design~\cite{MudarDCCBsizing}. During the faulted and post-fault states, it is, moreover, assumed that the infeed and demand setpoints, as well as the converter operating points, are equal to the pre-fault grid state. This entails that the generators located at infeeding nodes, such as (offshore) wind farms, ride through the network faults without the need for power curtailment, as long as sufficient network capacity is available. It also means that no automatic actions are taken to decrease the demand in onshore grids as a means to limit the fault impact, e.g. reserve activation or demand curtailment. 

In grid codes, the maximum loss of infeed is limited by the \textit{dimensioning incident}, stating that no installations may be connected if they have the potential to cause outages exceeding this limit, as they have the potential to cause intolerable frequency instabilities leading to black-outs. Consequently, it is assumed that limiting the loss of active power infeed, in worst-case fault conditions, is an accurate metric for the comparison of different protection designs~\cite{DCcont2017KUL}. In reality, the frequency instability of AC grids caused by a loss of active power infeed depends on several factors, such as the total grid inertia, the availability of frequency containment reserves, and the use of load curtailment~\cite{BrantlLjubljana2021,Dave2022Thesis}. Moreover, other AC grid aspects, i.e. transient stability and voltage stability, may be influenced by HVDC grid faults, but are less accurately represented by the loss of active power infeed. However, the loss of reactive power capability caused by DC grid faults can be modeled similarly to the possible loss of active power by evaluating the availability of converters during the faulted state of the grid. Nevertheless, in depth studies are required during the final phases of the HVDC grid design to assess the actual AC grid impacts of the active and reactive power losses. However, for the performed protection configuration shortlisting studies, the representation of this impact as momentary loss of infeed is sufficient. 

\subsection{Configuration selection process}
Using the graph-based modeling and fault impact evaluation method, different electrical energy hub configurations and protection designs can be generated and compared to create a shortlist of viable designs. The options are created by varying the connection points of HVDC cables, converters, and the mutual connections between the internal DC nodes in the electrical energy hub network model. The number of connected cables, converters, and the topology of the external grid are seen as inputs from the system designer that specify the overall constraints for the network and electrical energy hub for which the protection configuration is selected. 
Consequently, the shortlisting algorithm can be used to find the most effective configuration, from a protection standpoint, for a given energy hub size - represented by the number of cables and converters - and complexity or investment cost, represented by the DCCB count. This means that the number of included DCCBs can either be set as an input from the system designer, or the shortlisting can be performed for each DCCB count, allowing the effect of adding more DCCBs to be evaluated by comparing the selected configurations and fault impacts for different DCCB counts. 

The shortlisting methodology consists of an iterative process, starting with the creation of all possible configurations as graph networks and subsequently selecting those that achieve the lowest LoI in worst-case fault conditions in different power flow scenarios. The number of evaluated configurations depends on the electrical energy hub size and the allowed number of DCCBs. Cables in a single protection zone can be considered to be connected to the same node on the electrical energy hub from a protection zone point of view. This means that, although some switchgear other than DCCBs, i.e. not able to block DC faults, may be implemented between these connections, they are essentially seen as being connected to a single busbar. Consequently, the number of DC nodes in the electrical energy hub graph is equal to the number of protection zones achieved by the DCCB configuration. The total number of configurations is then set by the number of ways the converter and cable edges can be connected to the DC nodes, rising along with the converter and cable count. Moreover, the number of possible DC node configurations increases strongly if more DCCBs are allowed, further increasing the number of configurations that should be considered in the shortlisting process. This results in a very large set of $N$ possible configurations where $N$ can be expressed as
\begin{equation}
\label{eq:N_con}
    N = N_{con}\cdot(N_{DC,con})^{N_{cab}+N_{conv}},
\end{equation}

with $N_{con}$ the number of configurations that can be created with a certain number of DCCBs ($N_{DC}$) e.g. if three DCCBs are used, three configurations can be created, i.e a star, delta or line arrangement. $N_{DC,con}$ represents the number of DC nodes for a specific configuration and is limited to 
\begin{equation}
    N_{DC,con} \leq N_{CB} + 1,
\end{equation}
with $N_{CB}$ the number of DCCBs. As each converter and cable edge can be connected to any DC node, the number of possible configurations rises exponentially with the number of cables $N_{cab}$ and converters in the electrical energy hub $N_{conv}$ as shown in Eq.~\ref{eq:N_con}. 

The many possible configurations and the exponential increase in options as systems grow are clear indicators of the need to shortlist these design options. Consequently, the simplified network models are ideal for representing these large numbers of configurations, though the tractability of the algorithm is also limited when larger systems (i.e. with more than six cables or DCCBs) are evaluated. However, the initial solution size $N$ can be reduced, e.g. by eliminating configurations that do not utilize all available DC nodes and are thus identical to solutions with fewer breakers, or by eliminating options based on symmetry.

\subsection{Test case illustration}
\begin{figure}[htbp]
\centering
    \begin{subfigure}{.49\linewidth}
  \includegraphics[scale=1]{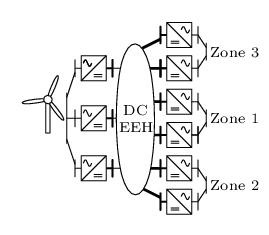}
  \caption{Medium test case}
  \label{fig:Medium_case}
\end{subfigure}\hfill 
\begin{subfigure}{.49\linewidth}
  \includegraphics[scale = 1]{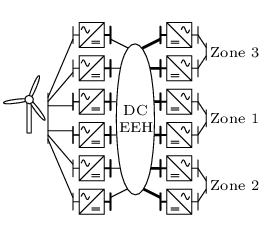}
  \caption{Large test case}
  \label{fig:Large_case}
\end{subfigure}
\caption{System-level diagram for medium and large electrical energy hub test case.}
\label{fig:Medium_Large_SLD}
\end{figure}

The shortlisting methodology is illustrated for three generic test cases, representing increasing energy hub sizes. Each of these cases has the same basic external network configuration, consisting of HVDC links between the electrical energy hub and three onshore grid zones. This configuration of the external HVDC grid is chosen to represent the context of several proposed electrical energy hub projects in Europe, which are expected to host interconnections to up to three different synchronous zones while integrating large capacities of offshore wind power into these onshore grids~\cite{EsbjergCooperation,DanishWebsite, Bornholm, Elia2023Website}. As these projects are expected to increase in size in the future, the chosen test cases have both an increasing capacity of connected cables and offshore wind capacity at the electrical energy hub. Tab.~\ref{tab:testcases} lists the number of converters on the electrical energy hub and cable connections to each onshore zone (per pole) for each test case. The system-level diagram for the small test case is depicted in Fig.~\ref{fig:small_case_SLD} and introduced for the medium and large test cases in Fig.~\ref{fig:Medium_case} and~\ref{fig:Large_case}. 
\begin{table}[htbp]
    \centering
    \caption{Parameters (per pole) for the considered test cases.}
    \label{tab:testcases}
    \begin{tabular}{lcccc}
    \toprule
        Test Case & Internal & \multicolumn{3}{c}{Connections to} \\  & Converters & Zone 1& Zone 2 & Zone 3\\ \midrule
        Small & 2 & 2 & 1 & 1\\
        Medium & 3 & 2 & 2 & 2\\
        Large & 6 & 2 & 2 & 2\\ \bottomrule
    \end{tabular}
\end{table}

\begin{figure}[htbp]
    \centering
    \includegraphics[width=\linewidth]{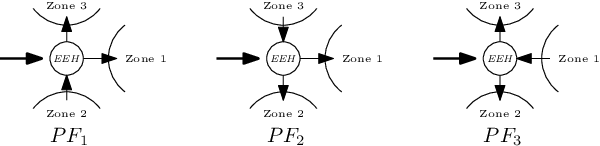}
    \caption{Example power flow scenarios used to illustrate the shortlisting process.}
    \label{fig:PowerFlowScenarios}
\end{figure}

The shortlisting process filters electrical energy hub configurations by evaluating the fault impact in different power flow scenarios. In this paper, the method is illustrated for the example test cases using three example power flow scenarios ($PF_1$-$PF_3$). The converters on the electrical energy hub are assumed to feed power into the HVDC grid at their rated capacity, representing high wind infeed scenarios. The remaining power flows are then varied by changing which onshore zone acts as a net exporter, leading to the three power flow scenarios as depicted by Fig.~\ref{fig:PowerFlowScenarios}. In the small, and medium test cases, these power flow scenarios accurately describe the different possible conditions for high-impact faults in high wind infeed conditions. However, in the large test case, only one power flow scenario is considered as the wind infeed from the electrical energy hub is enough to fully load the system and no external export zone is required.

\begin{figure}[htbp]
    \centering
    \includegraphics[scale=1]{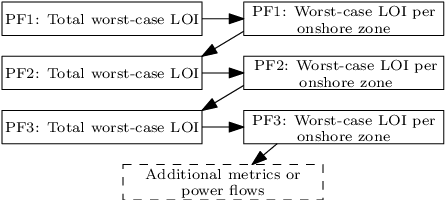}
    \caption{Filtering steps in the considered example of the shortlisting methodology.}
    \label{fig:filtering_flowchart}
\end{figure}

In each considered power flow scenario, electrical energy hub configurations are selected that achieve the lowest total worst-case loss of power transfer, as well as the lowest impact per AC zone. Given the three considered power flow scenarios, this results in six basic filtering steps in which the number of applicable electrical energy hub configurations can be significantly reduced. However, further reductions can be made if necessary by filtering the configuration options based on additional metrics, rather than only the worst-case loss of power transfer. Fig.~\ref{fig:filtering_flowchart} illustrates these filtering steps. 

%% file: 3-Results.tex
\section{Results}
\label{sec:results}
\subsection{Filtering outcomes}
\label{sec:filtering outcomes}
The shortlisting process is demonstrated in this section for the three considered test cases. For each case, configurations with up to five DCCBs are considered, resulting in a large number of evaluated configurations, especially for the large test case.

Using the shortlisting methodology, the possible configurations can be drastically reduced, retaining only the design options that achieve minimal worst-case fault impacts in the considered power flow scenarios, i.e. \textit{lowest impact configurations}. Only these lowest impact configurations are then evaluated in subsequent filtering steps using different fault impact metrics and power flow scenarios. After considering all metrics and power flows, the final set of lowest impact configurations represent the most effective protection configurations for the considered HVDC switching station.

\begin{figure}[htbp]
\begin{subfigure}{\linewidth}
\centering
  \includegraphics[scale=1]{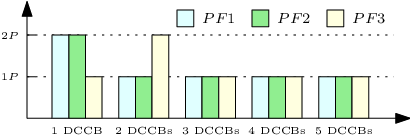}
  \caption{Small test case}
  \label{fig:Results_small}
\end{subfigure}\hfill 

\medskip 
\begin{subfigure}{\linewidth}
\centering
  \includegraphics[scale = 1]{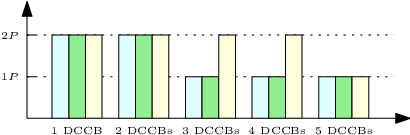}
  \caption{Medium test case}
  \label{fig:Results_medium}
\end{subfigure}\hfill 

\medskip 
\begin{subfigure}{\linewidth}
\centering
  \includegraphics[scale = 1]{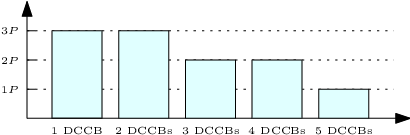}
  \caption{Large test case}
  \label{fig:Results_large}
\end{subfigure}\hfill 
\caption{Total worst-case loss of power transfer in each considered power flow scenario in the example electrical energy hub test cases for the selected lowest impact configurations.}
\label{fig:ImpactResultsAll}
\end{figure}

Fig.~\ref{fig:ImpactResultsAll} shows the worst-case loss of power transfer to all zones in the lowest impact configurations in each power flow scenario. This figure shows that, to reduce the fault impact to $1P$ in all power flow conditions, the small test case requires at least three DCCBs while the medium and large cases need at least five. However, if fewer DCCBs are used, it is possible in some cases to reduce the impact only for specific power flow scenarios. This result highlights the influence of power flows on the performance of the protection system. Consequently, the ability to separately consider and prioritize power flows is an advantage of the proposed shortlisting method, compared to system-level protection strategy-based choices.

\subsection{Additional filtering metrics}
\begin{figure}[htbp]
\begin{subfigure}{\linewidth}
\centering
  \includegraphics[scale=1]{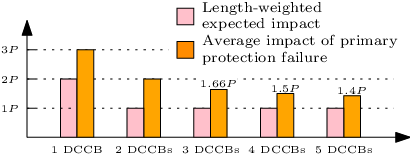}
  \caption{Small test case}
  \label{fig:Additional_Results_small}
\end{subfigure}\hfill 

\medskip 
\begin{subfigure}{\linewidth}
\centering
  \includegraphics[scale = 1]{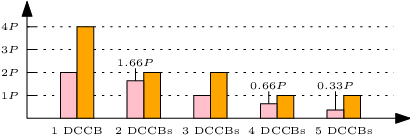}
  \caption{Medium test case}
  \label{fig:Additional_Results_medium}
\end{subfigure}\hfill 

\medskip 
\begin{subfigure}{\linewidth}
\centering
  \includegraphics[scale = 1]{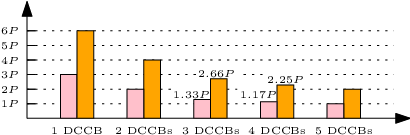}
  \caption{Large test case}
  \label{fig:Additional_Results_large}
\end{subfigure}\hfill 
\caption{Length-weighted expected impact and average impact of primary protection failure in the considered electrical energy hub test cases for the selected protection configurations.}
\label{fig:Additional_Impact}
\end{figure}

Besides the worst-case loss of power transfer, any metric that relies on the connectivity of the HVDC grid before and during the faulted state can be applied in the proposed methodology. These metrics can either be used in place of the worst-case impact metric or to carry out additional filtering steps after the initial selection of lowest impact configurations. Examples of additional metrics are the availability of reactive power during faults, i.e. the continuous operation of offshore and onshore converters, or the preferred configuration if specific network expansions are expected in the future. Here, two other additional metrics are introduced: \textit{length-weighted expected fault impact} and \textit{average impact of primary protection failure}. The results for these metrics applied to the considered test cases for power flow scenario $PF_1$ are shown in Fig.~\ref{fig:Additional_Impact}.

The length-weighted expected fault impact metric ($P_{avg}$) represents the expected impact of faults rather than the worst-case. Since it could be estimated that the failure rate of an HVDC cable is approximately proportional to the length, the expected fault impact can simply be evaluated by averaging the impact of every cable fault ($c$), weighted by the length of each cable ($L_c$) compared to the total length of all cables in the system ($L_{tot}$) as shown in Eq.~\ref{eq:avg_impact}.
\begin{align}
\label{eq:avg_impact}
    P_{avg} = \sum_c\left(\frac{L_c}{L_{tot}}\cdot\sum_G P_{PreFault,G} - P_{Fault,G,c}\right)
\end{align}

In the considered test cases, the length-weighted expected impact metric is applied to the selected topologies after the initial filtering steps (which evaluate the maximum impact) with cable lengths assumed as 100~km, 700~km and 250~km for connections to Zone 1, 2 and 3 respectively. 

The second additional metric is the average loss of active power transfer in case of primary protection failure ($P_{bu}$), i.e. in case the \textit{back-up} protection needs to operate. The failure of a primary DCCB to interrupt a fault is assumed. This can be caused by a malfunction of the breaker itself or the controlling protection algorithms. It is assumed here that the fault is, in this case, interrupted by adjacent DCCBs or AC circuit breakers, causing a larger portion of the HVDC grid to be de-energized. The impact of primary protection failure is calculated as the average loss of power transfer caused by the worst-case cable fault ($c$) for every primary breaker failure ($PB$):
\begin{align}
    P_{bu} = \frac{1}{N_{CB}}\cdot \sum_{PB}\left(\max_c\left(\sum_G P_{PreFault,G}-P_{Fault,G,c,PB}\right)\right)
\end{align}

\begin{figure*}[htbp]
\begin{subfigure}{0.32\textwidth}
  \includegraphics[width=\linewidth,clip=true,trim=11mm 1mm 11mm 6.4mm]{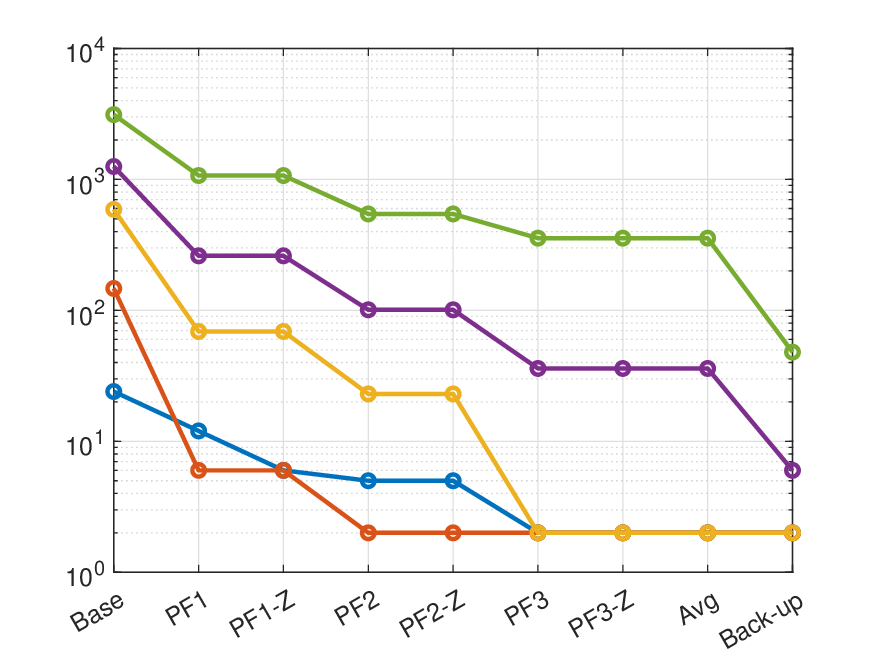}
  \caption{Small test case}
  \label{fig:small_filter_steps}
\end{subfigure}\hfill 
\begin{subfigure}{.32\textwidth}
  \includegraphics[width=\linewidth,clip=true,trim=11mm 1mm 11mm 6.4mm]{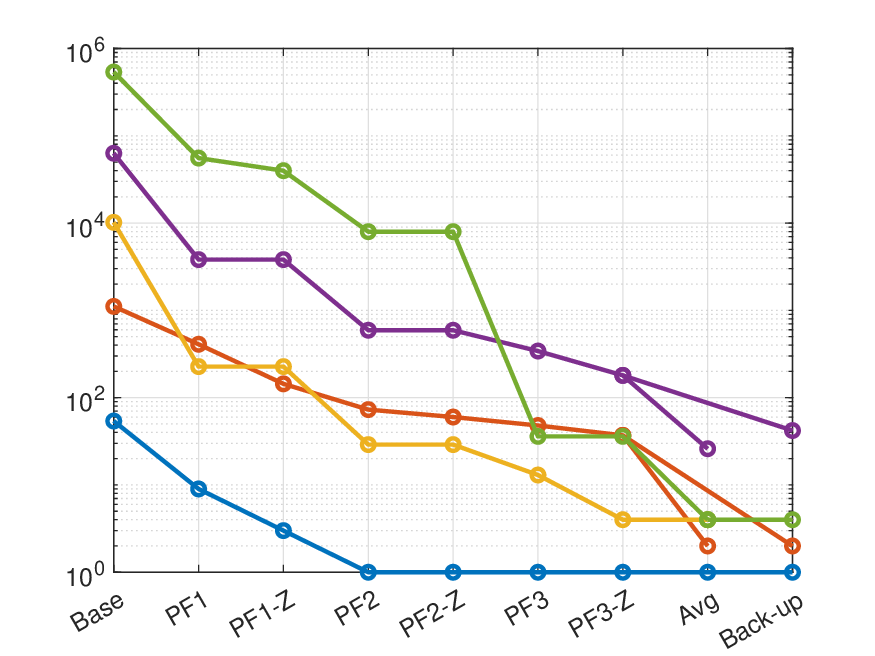}
  \caption{Medium test case}
  \label{fig:medium_filter_steps}
\end{subfigure}\hfill
\begin{subfigure}{.32\textwidth}
  \includegraphics[width=\linewidth,clip=true,trim=11mm 1mm 11mm 6.4mm]{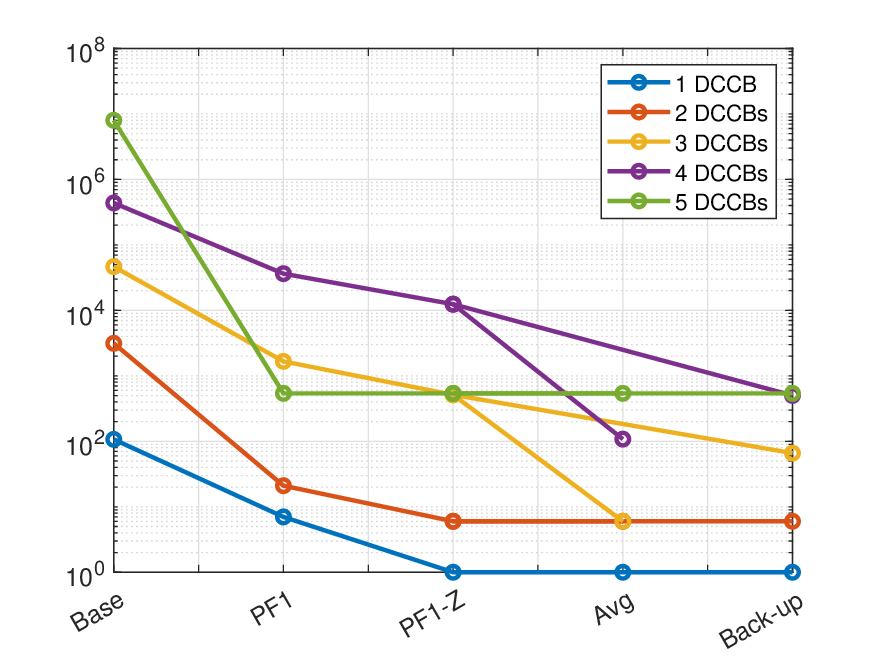}
  \caption{Large test case}
  \label{fig:large_filter_steps}
\end{subfigure}
\caption{Number of possible protection configurations in the considered electrical energy hub test cases at different stages of the filtering process. The additional metric evaluations are executed in parallel and drawn as separate lines if the results differ.}
\label{fig:filtering_summary}
\end{figure*}
\subsection{Discussion of results}
The protection configuration shortlisting methodology has been applied to three test cases that represent example electrical energy hub networks of increasing sizes. The need for an efficient filtering process follows from the number of possible configurations, which becomes increasingly large - even after preemptively eliminating illogical cases - especially in electrical energy hubs with many connections and DCCBs. In cases with many initial options, many configurations remain after the initial filtering steps. However, by using additional metrics for protection design evaluation in the filtering process, the lowest impact configurations can be further reduced until only configurations that function identically from a protection perspective remain. 
The filtering process and the reduction of possible protection configurations are summarized in Fig.~\ref{fig:filtering_summary} where the number of lowest impact configurations is shown (on a logarithmic scale) for each filtering step.

Besides being a method for efficient electrical energy hub design, the shortlisting methodology allows system designers to evaluate how many DCCBs are required in the system to achieve a certain maximum fault impact, while ensuring that DCCBs are efficiently placed. The additional metrics can also be applied to further differentiate between different DCCB counts. The total worst-case fault impact results, shown in Fig.~\ref{fig:ImpactResultsAll} show that, in the small test case, no difference in impact is achieved by increasing the DCCB  count beyond three. Similarly, in the medium and large cases, there is no reduction in worst-case impact if two or four DCCBs are used compared to the selected configurations with one fewer breaker. However, in these cases, increasing the DCCB count reduces the length-weighted average impact or the average impact of primary protection failure, as shown by Fig.~\ref{fig:Additional_Impact}. Thus, different DCCB counts can be evaluated based on these metrics.

\begin{figure}[htbp]
\begin{subfigure}{0.32\linewidth}
  \includegraphics[width=\linewidth]{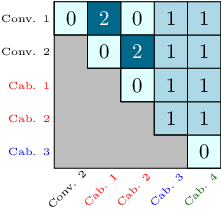}
  \caption{1 DCCB}
  \label{fig:SB1_Heat_Map}
\end{subfigure}\hfill 
\begin{subfigure}{.32\linewidth}
  \includegraphics[width=\linewidth]{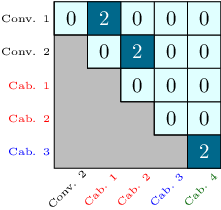}
  \caption{2 DCCBs}
  \label{fig:SB2_Heat_Map}
\end{subfigure}\hfill
\begin{subfigure}{.32\linewidth}
  \includegraphics[width=\linewidth]{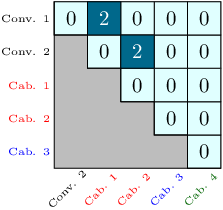}
  \caption{3 DCCBs}
  \label{fig:SB3_Heat_Map}
\end{subfigure}
\caption{Number of times elements are located in the same protection zone in the lowest impact configurations in the small test case. Label colors represent connected AC zones.}
\label{fig:Heat_Maps}
\end{figure}

By evaluating these different levels of protection represented by increasing DCCB counts, the priorities for protection designs can be studied. 
Fig.~\ref{fig:Heat_Maps} illustrates these principles by representing the lowest impact configurations for the small test case as heat maps. These heat maps show how often elements (converters and cables) are connected in the same protection zones. Fig.~\ref{fig:SB1_Heat_Map} shows that, if only one DCCB is implemented, it is used to separate each converter on the electrical energy hub, as well as the two cables connecting to AC Zone 1 into different zones. Figures~\ref{fig:SB2_Heat_Map} and \ref{fig:SB3_Heat_Map} show that this principle is maintained even if more DCCBs are included. The separation of these power flow paths from the energy hub converters to the AC zone with two connected cables shows both the importance of minimizing the fault impact per synchronous zone, as well as the prioritization of the first two considered power flows, which both feed power into AC~Zone~1. Fig.~\ref{fig:SB1_Heat_Map} also shows that Cables~3 and 4, connecting to AC Zones 2 and 3 respectively are divided over the two protection zones created by the one DCCB, thus explaining why, in the scenario where all power flows to these two zones, $PF_3$, the impact can be minimized to $1P$. 
It can be seen in Figures~\ref{fig:Results_small} and~\ref{fig:SB2_Heat_Map} that putting Cables~3 and 4 in one separate protection zone by using an extra DCCB reduces the worst-case fault impact in $PF_1$ and $PF_2$ while it is increased in $PF_3$. Fig.~\ref{fig:SB3_Heat_Map} shows that, only if all cables connecting to onshore AC zones are separated into different protection zones, the fault impact can be minimized to $1P$ in every power flow scenario.

Though it is evident that impacts can be minimized by connecting each component in a different protection zone if several DCCBs are used, the results in Fig.~\ref{fig:Heat_Maps} also show less obvious principles regarding protection design with fewer DCCBs. A similar evaluation can be made for the medium and large test cases.  This is illustrated in Tab.~\ref{tab:design_rules} which lists, for each test case, the proportion of lowest impact cases\footnote{After the full filtering process, including the average expected impact additional metric.} for all considered breaker counts to which the following design principles apply:
\begin{enumerate}
    \item Cables to the same AC zone are separated in different protection zones
    \item All converters in the EEH are separated in different protection zones.
    \item $PF_1$ is prioritized by connecting cables to AC zone 1 in the same protection zone as EEH converters
    \item $PF_1$ is prioritized by separating connections to AC~zone~1 from other cables
\end{enumerate}

The results in Tab.~\ref{tab:design_rules} highlight the key principles for the considered test cases. It is shown that, in the lowest impact configurations, connections to the same AC zone are always separated and that converters in the electrical energy hub are also often connected in separate protection zones. Additionally, the power flow to AC zone 1 is prioritized in different ways. In the small and large test cases, connections to AC zone 1 are always placed in the same protection zone as converters. In the medium test case, connections to AC zone 1 are usually separated from other connections. While these analyses provide insights into the design principles for the considered test cases, future work considering various additional test cases could lead to more generally applicable protection design rules.
\begin{table}[htbp]
    \caption{Occurrence rate of proposed design principles in lowest impact configurations after shortlisting.}
    \label{tab:design_rules}
    \centering
    \begin{tabular}{cccc}
    \toprule
        Principle &  Small & Medium & Large\\ \midrule
        (1) & 100\% & 100\% & 100\% \\
        (2) & 100\% & 97.3\% & 81.7\%\\
        (3) & 100\% & 51.4\% & 100\%\\
        (4) & 66.7\% & 91.9\% & 81.7\%\\ \bottomrule
    \end{tabular}
\end{table}

\subsection{Evaluation of fully selective configurations}
In the performed case studies, protection configurations with up to five DCCBs are considered. This allows each cable, in the largest considered test cases, to be connected in a separate protection zone with respect to other cable connections. Although the implementations of traditional protection strategies are not uniquely defined in an electrical energy hub or an MTDC grid with many centralized connections, example configurations resembling these strategies can be compared with the configurations that are selected by the full shortlisting process. Based on the conclusions from~\cite{SuperGridInstitute}, a possible implementation of the fully selective strategy can be considered in which a DCCB is included on each cable or line connection on the electrical energy hub, but not at the converter connections. By implementing this specific configuration according to the network modeling methodology used in the shortlisting process, the results for this specific implementation in the small, medium and large test cases are calculated. The achieved impacts with this configuration and the required number of DCCBs are listed for the considered power flow scenarios and additional metrics in Tab.~\ref{tab:FS_reference results}. These results show that the typical fully selective strategy maximally limits the maximum worst case and average expected fault impact but is subject to very high fault impacts in case of primary protection failure. Moreover, if these results are compared to the impacts achieved in the configurations selected by the shortlisting method, as displayed in Fig.~\ref{fig:ImpactResultsAll} and~\ref{fig:Additional_Impact}, it becomes clear that the same worst case and average fault impacts can be achieved with one fewer DCCB, while the impact of primary protection failure is also reduced compared to the fully selective implementation. This demonstrates the importance of considering the protection configuration as proposed in this paper, instead of implementing fixed protection strategies. 
\begin{table}[htbp]
    \centering
    \caption{Fault impacts in the configuration from~\cite{SuperGridInstitute} compared to shortlisting results (\textbf{bold}) where different.}
    \label{tab:FS_reference results}
    \begin{tabular}{lcccccc}
    \toprule
        Test case&DCCBs & $PF_1$ & $PF_2$ & $PF_3$ & AVG & BU \\ \midrule
        Small & 4 (\textbf{3}) & $1P$ & $1P$ & $1P$ & $1P$ & $3P$ (\textbf{1.66\textit{P}}) \\
        Medium & 6 (\textbf{5}) & $1P$ & $1P$ & $1P$ & $0.33P$ & $4P$ (\textbf{1\textit{P}}) \\
        Large & 6 (\textbf{5}) & $1P$ & $1P$ & $1P$ & $1P$ & $6P$ (\textbf{2\textit{P}}) \\ \bottomrule
    \end{tabular}   
\end{table}

In future studies, the protection configuration shortlisting methodology will be used for further evaluations of proposed HVDC switching station designs.

%% file: 4-Conclusion.tex
\section{Conclusion}
The presented shortlisting approach for protection configurations in electrical energy hubs offers a new method for HVDC protection system design. This method integrates the configuration of DCCBs and the connection of HVDC converters and cables into the system design during early development stages of electrical energy hubs or HVDC switching stations. Reduced-order models are used to evaluate large quantities of protection configurations, including possible designs that are not described by typical protection strategies. Consequently, a more open approach to the protection design is adopted by considering every possible protection configuration option. These options are assessed based on several fault impact metrics using a simplified fault impact calculation method that does not require a detailed analysis of the HVDC or AC grids. 

The fault impact results generated during the shortlisting process allow system developers to decide on the required investment into HVDC protection equipment by evaluating impacts in different scenarios. Since the filtering method selects only the configurations that achieve the lowest impacts with a given DCCB count, it is also ensured that equipment is efficiently used for the considered fault scenarios.

When applied to three example electrical energy hub test cases, it is shown that the shortlisting methodology facilitates a strong reduction in the number of possible protection configurations through the iterative filtering process. 

The selected lowest impact configurations at the end of the filtering process show that the protection configuration should generally place connections to one AC grid in different protection zones and maximally separate the most relevant power flow paths. Moreover, the comparison of a typical implementation of the fully selective strategy to the configurations selected by the shortlisting methodology shows that the proposed approach leads to improved protection designs that, with fewer DCCBs than the typically considered implementation of the fully selective strategy, equally limit the worst-case and average impact of faults, while resulting in lower fault impacts in case of primary protection failure.

%% file: Main.bbl
\begin{thebibliography}{10}

\bibitem{OstendDeclaration}
{Governments of the countries in the North Sea area}, ``Ostend declaration of energy ministers on the north seas as europe's green power plant: Delivering cross-border projects and anchoring the renewable offshore industry in europe.'' [Online] Available: https://northseasummit.fedbook.prd.excom.fgov.be/en/ostend-declaration. (accessed November 17, 2023), 2023.

\bibitem{bookDirk}
D.~Van~Hertem, O.~Bellmunt, and J.~Liang, {\em HVDC grids for transmission of electrical energy: Offshore grids and a future supergrid}.
\newblock Wiley-IEEE Press, 03 2016.

\bibitem{Elia2023Website}
{Elia}, ``Princess elisabeth island.'' [Online] Available: https://www.elia.be/en/infrastructure-and-projects/infrastructure-projects/princess-elisabeth-island. (accessed November 17, 2023).

\bibitem{DanishWebsite}
{\O}rsted and ATP, ``North sea energy island.'' [Online] Available: northseaenergyisland.dk. https://northseaenergyisland.dk/en. (accessed November 21, 2023).

\bibitem{Bornholm}
{Danish Energy Agency}, ``Bornholm energy island.'' [Online] Available: https://ens.dk/en/our-responsibilities/onshore-wind-power/bornholm-energy-island. (accessed February 14, 2024).

\bibitem{EsbjergCooperation}
50Hertz, Ampiron, Elia, Energinet, Gasunie, and Tennet, ``The esbjerg cooperation, transforming the north sea into europe's green power plant,'' 2023.

\bibitem{Dirk2010DCgrids}
D.~Van~Hertem and M.~Ghandhari, ``Multi-terminal vsc hvdc for the european supergrid: Obstacles,'' {\em Renewable and Sustainable Energy Reviews}, vol.~14, no.~9, pp.~3156--3163, 2010.

\bibitem{leterme2019designing}
W.~Leterme, I.~Jahn, P.~Ruffing, K.~Sharifabadi, and D.~Van~Hertem, ``Designing for high-voltage dc grid protection: Fault clearing strategies and protection algorithms,'' {\em IEEE Power and Energy Magazine}, vol.~17, no.~3, pp.~73--81, 2019.

\bibitem{PROMOTioN_D4_2}
{PROMOTioN Project}, ``{D4.2 – Broad comparison of fault clearing strategies for DC grids},'' 2017.

\bibitem{subedi2021review}
S.~Subedi, M.~Rauniyar, S.~Ishaq, T.~Hansen, R.~Tonkoski, M.~Shirazi, R.~Wies, and P.~Cicilio, ``Review of methods to accelerate electromagnetic transient simulation of power systems,'' {\em IEEE Access}, vol.~9, pp.~89714--89731, 2021.

\bibitem{NSWPH_II}
A.~Zama, A.~Bertinato, P.~Torwelle, W.~L. Garcia, J.~V. Doorn, J.~P. Kjærgaard, F.~Kryezi, and A.~M. Lindefelt, ``North sea wind power hub feasibility study: Methodology for protection design (part ii),'' in {\em 2022 International Conference on Renewable Energies and Smart Technologies (REST)}, vol.~I, pp.~1--5, 2022.

\bibitem{SuperGridInstitute}
{SuperGrid Institute}, ``Nswph validation technical requirements: Sow a final feasibility report,'' 2022.

\bibitem{PROMOTioN_D4_7}
{PROMOTioN Project}, ``{D4.7 - Preparation of cost-benefit analysis from a protection point of view},'' 2020.

\bibitem{DCcont2017KUL}
M.~Abedrabbo, M.~Wang, P.~Tielens, F.~Z. Dejene, W.~Leterme, J.~Beerten, and D.~{Van Hertem}, ``{Impact of DC grid contingencies on AC system stability},'' in {\em Proc. IET ACDC 2017}, (Manchester, UK), 2017.
\newblock 7~pages.

\bibitem{BrantlLjubljana2021}
C.~Brantl, M.~Knechtges, P.~Düllmann, C.~Meier, and A.~Moser, ``Requirements on offshore hvdc grid protection: Interaction with ac system inertia and fast frequency support,'' in {\em 41. CIGRE International Symposium}, (Ljubljana, SI), 2021.

\bibitem{Dave2022Thesis}
J.~Dave, {\em DC grid protection aware planning of offshore HVDC grids}.
\newblock PhD thesis, KU Leuven, 2022.

\bibitem{Preventive_Decoupling_paper}
P.~Düllmann, C.~Klein, P.~Winter, H.~Köhler, M.~Steglich, J.~Teuwsen, and A.~Moser, ``Preventive dc-side decoupling: a control and operation concept to limit the impact of dc faults in offshore multi-terminal hvdc systems,'' {\em 19th International Conference on AC and DC Power Transmission (ACDC 2023)}, pp.~30--37, 2023.

\bibitem{SGI_non_selective}
A.~Bertinato, J.~Gonzales, D.~Loume, C.~Creusot, B.~Luscan, and B.~Raison, ``Development of a protection strategy for future dc networks based on low-speed dc circuit breakers,'' in {\em Proc. CIGRE Session}, (Paris, FR), 2018.

\bibitem{DantasProgressiveIsolation2018}
R.~Dantas, J.~Liang, C.~E. Ugalde-Loo, A.~Adamczyk, C.~Barker, and R.~Whitehouse, ``Progressive fault isolation and grid restoration strategy for mtdc networks,'' {\em IEEE Transactions on Power Delivery}, vol.~33, no.~2, pp.~909--918, 2018.

\bibitem{MudarDCCBsizing}
M.~Abedrabbo, W.~Leterme, and D.~Van~Hertem, ``Systematic approach to hvdc circuit breaker sizing,'' {\em IEEE Transactions on Power Delivery}, vol.~35, no.~1, pp.~288--300, 2020.

\end{thebibliography}
